\documentclass{article}

\PassOptionsToPackage{numbers, compress}{natbib}
\usepackage[final]{neurips_2024}
\usepackage[utf8]{inputenc} % allow utf-8 input
\usepackage[T1]{fontenc}  % use 8-bit T1 fonts
\PassOptionsToPackage{hyphens}{url}\usepackage{hyperref}   % hyperlinks
\usepackage{url}    % simple URL typesetting
\usepackage{booktabs}   % professional-quality tables
\usepackage{amsfonts}   % blackboard math symbols
\usepackage{nicefrac}   % compac ypography
\usepackage{xcolor}   % colors

\title{\stepcounter{footnote}``{\em I Am the One and Only, Your Cyber BFF}'':\thanks{The title is inspired by a response that a Reddit user received when using the Pi chatbot~\cite{PiClaimsToBeChatGPT}.} Understanding the Impact of GenAI Requires Understanding the Impact of Anthropomorphic AI}

\author{
Myra Cheng \\
Stanford University 
\And
Alicia DeVrio\\
Carnegie Mellon University
\And
Lisa Egede \\
Carnegie Mellon University
\And
Su Lin Blodgett* \\
Microsoft Research
\And
Alexandra Olteanu* \\
Microsoft Research
}

\begin{document}
\maketitle

\setcounter{footnote}{0} 

\section{Anthropomorphic AI System Behaviors Are Prevalent Yet Understudied}

%%%Introducing the phenomenon and broadly situating/linking it to understanding the impacts of AI systems
In his 1985 lecture, Edsger Dijkstra lamented that anthropomorphism was rampant in computing science, with many of his colleagues perhaps not realizing how pernicious it was, and that ``{\em [i]t is not only the [computing] industry that suffers, so does the science}''~\cite{dijkstra1985anthropomorphism}.
Indeed, {\em anthropomorphism}---or the attribution of human traits to non-human entities---in how we talk about computing systems shapes how people understand and interact with AI and other computing systems \citep{cheng-etal-2024-anthroscore,nass1994computers,reeves1996media}, and is thus at the core of understanding the impacts of these systems on individuals, communities, and society.\looseness=-1

%%%Not only how we talk about systems, but also system behavior. Increasing prevalence of anthropomorphic behaviors, either unintentional or intentional
But it is not only how we talk about computing systems. 
Many state-of-the-art generative AI (GenAI) systems are increasingly prone to anthropomorphic behaviors \cite[e.g.,][]{abercrombie2023mirages,agnew2024illusion,chan2023harms,gabriel2024ethics}---i.e., to generating outputs that are {\em perceived} to be human-like---either by design~\cite{mcilroy2022mimetic,park2022social,park2023generative} or as a by-product of how they are built, trained, or fine-tuned~\cite{bender2021dangers,tjuatja2024llms}. 
For instance, LLM-based systems have been noted to output text claiming to have tried pizza~\cite{pizzatweet}, 
to have fallen in love with someone \cite{roose2023conversation}, to be human or even better than humans \cite{decosmo2022google}, and to have human-like life experiences \cite{fiesler2024ai}. 
Such {\em anthropomorphic systems}\footnote{We deliberately use the terms {\em anthropomorphic AI}, {\em anthropomorphic systems} or {\em anthropomorphic system behaviors}---systems and system outputs that are {\em perceived} to be human-like---instead of {\em agentic systems}~\cite{chan2023harms,shavit2023practices} or {\em human-like AI}~\cite{brynjolfsson2023turing} to emphasize that these systems are rather believed to be human-like or have human-like characteristics based on {\em perceptions}; we thus try to steer clear of inadvertently suggesting that AI systems are human or have human-like agency or consciousness. That is, a stone being perceived as human-like does not necessarily imply the stone is human. We similarly avoid ambiguous, speculative, or relative terms whose meanings are likely to change across contexts or over time, such as {\em advanced AI}~\cite{gabriel2024ethics} (a term used as early as the 80s) or {\em emergent properties}~\cite{rogers2024position}. We instead focus on developers' stated design goals---what systems are intended to do---and in what ways AI outputs might be perceived as human-like, rather than on what systems can or cannot do.  \looseness=-1} 
range from conversational assistants~\cite[e.g.,][]{abercrombie2021alexa,shanahan2023role} to avatars and chatbots designed as a stand-in for friends, companions, or romantic partners~\cite[e.g.,][]{AI-romantic-partner,brandtzaeg2022my,laestadius2022too,ruiz2024marshable}, and AI-generated media designed to portray people~\cite[e.g.,][]{rosner2021ethics,vaccari2020deepfakes}, among a fast-growing number of applications~\cite[e.g.,][]{agnew2024illusion,mcilroy2022mimetic,ChatGPT-human}.\looseness=-1

While scholars have increasingly raised concerns about a range of possible negative impacts from anthropomorphic AI systems
\cite[e.g.,][]{abercrombie2023mirages,bender2021dangers,friedman1992human,ibrahim2024characterizing,maeda2024human}, anthropomorphism in AI development, deployment, and use remains vastly overlooked, understudied, and underspecified. 
Without making hard-and-fast claims about the merits (or the lack thereof) of anthropomorphic systems or system behaviors, we believe we need to do more to develop the know-how and tools to better tackle anthropomorphic behavior, including measuring and mitigating such system behaviors when they are considered undesirable.
Doing so is critical because---among many other concerns---having AI systems generating content claiming to have e.g., feelings, understanding, free will, or an underlying sense of self may erode people’s sense of agency~\cite{friedman1992human}, with the result that people might end up attributing moral responsibility to systems~\cite{friedman1992human,friedman2007human}, overestimating system capabilities \cite{friedman2007human,Watson2019-py}, or overrelying on these systems even when incorrect~\cite{abercrombie2023mirages,kim2024m,Zarouali2021-gy}.\looseness=-1

In this brief perspective, we argue that as GenAI systems are increasingly anthropomorphic, {\em we cannot thoroughly map the landscape of possible social impacts of GenAI without mapping the social impacts of anthropomorphic AI}. 
We believe that drawing attention to anthropomorphic AI systems helps foreground particular risks---e.g., that people may develop emotional dependence on AI systems~\cite{laestadius2022too}, that systems may be used to simulate the likeness of an individual or a group without consent~\cite{bariach2024towards,whitney2024real,widder2022limits}, or the instrumentalization or dehumanization of people~\cite{aizenberg2020designing,erscoi2023pygmalion,van2024artificial}---that might otherwise be less salient or obscured by a focus on more widely recognized or understood risks such as those related to fairness harms~\cite{bennett2020point,olteanu2023responsible,weinberg2022rethinking}.\looseness=-1

\section{A Call to Action}

The foregrounding of (un)fair system behaviors in recent years \cite{barocas-hardt-narayanan} is nevertheless instructive, as it illustrates the dividends we have gotten from making fairness a critical concern about AI systems and their behaviors: better conceptual clarity about the ways in which systems can be unfair or unjust \cite[e.g.,][]{benjamin2019race,crawford2017neurips}, a richer set of measurement and mitigation practices and tools \cite[e.g.,][]{blodgett-etal-2021-stereotyping,jacobs_measurement_2021}, and deeper discussions and interrogations of underlying assumptions and trade-offs \cite[e.g.,][]{hoffmann2019fairness,jakesch2022different,keyes2019mulching}. 

We argue that a focus on anthropomorphic systems and their behaviors will similarly encourage a deeper interrogation of the ways in which systems are anthropomorphic, the practices that lead to anthropomorphic systems, and the assumptions surrounding the design, deployment, evaluation, and use of these systems, and is thus likely to yield similar benefits.\looseness=-1

\noindent{\bf We need more conceptual clarity around what constitute anthropomorphic behaviors.} 
Investigating anthropomorphic AI systems and their behaviors can, however, be tricky because language, as with other targets of GenAI systems, is itself innately human, has long been produced by and for humans, and is often also about humans. 
This can make it hard to specify appropriate alternative (less human-like) behaviors, and risks, for instance, reifying harmful notions of what---and whose---language is considered more or less human \cite{wynter2003unsettling}.

Understanding what exactly constitute anthropomorphic behaviors, and thus in what ways system behaviors are anthropomorphic, is nonetheless necessary to measure and determine which types of behaviors should be mitigated and how, and which behaviors are perhaps desirable (if any at all). 
This requires unpacking the wide range of dynamics and varieties in system outputs that are potentially anthropomorphic. 
While a system output including expressions of politeness like ``{\em you're welcome}'' and ``{\em please}'' (known to contribute to anthropomorphism~\cite[e.g.,][]{fink2012anthropomorphism}) might in some deployment settings be deemed desirable, 
system outputs that include suggestions that a system has a human-like identity or self-awareness---such as through expressions of self as human (``{\em I think I am human at my core}''~\cite{sentientGoogle}) or through comparisons with humans and non-humans (``{\em[language use] is what makes us different than other animals}''~\cite{sentientGoogle})---or that include claims of physical experiences---such as sensory experiences (``{\em when I eat pizza}''~\cite{pizzatweet}) or human life history (``{\em I have a child}''~\cite{haschildtweet})---might not be desirable.\looseness=-1

\noindent{\bf We need deeper examinations of both possible mitigation strategies and their effectiveness in reducing anthropomorphism and attendant negative impacts.} 
Intervening on anthropomorphic behaviors can also be tricky as people may have different or inconsistent conceptualizations of what is or is not human-like~\cite{abercrombie2023mirages,heyselaar2023casa,lang2013computers}, and sometimes the same system behavior can be perceived differently in different contexts; for example, expressions of uncertainty in system outputs may sometimes be associated with human-like equivocation and other times with objectivity (and thus with more machine-likeness~\cite[e.g.,][]{quintanar1982interactive}). 
Interventions intended to mitigate anthropomorphic system behaviors can thus fail or even heighten anthropomorphism (and attendant negative impacts) when applied or operationalized uncritically. 
For instance, a commonly recommended intervention is including in the AI system's output a disclosure that the output is generated by an AI system~\cite[e.g.,][]{el2024transparent,google-disclosure,mozafari2020chatbot,van2024understanding}. 
How to operationalize such interventions in practice and whether they can be effective alone might not always be clear.  
For instance, while the example ``{\em [f]or an AI like me, happiness is not the same as for a human like you}''~\cite{roach2023want} includes a disclosure, it may still suggest a sense of identity and ability to self-assess (common human traits). 
 
\noindent{\bf We need to interrogate the assumptions and practices that produce anthropomorphic AI systems.}
Understanding and mitigating the impacts of anthropomorphic systems also requires us to interrogate how the assumptions and practices that underlie the development and deployment of these systems may lead (purposefully or otherwise) to anthropomorphic system behaviors.
For example, current approaches to collecting human preferences about system behavior (e.g., RLHF) do not consider the differences between what may be appropriate for a response from a human versus from an AI system; a statement that seems friendly or genuine from a human speaker can be undesirable if it arises from an AI system since the latter lacks meaningful commitment or intent behind the statement, thus rendering the statement hollow and deceptive \cite{winograd1986understanding}. 
Doing so will also help provide a more robust foundation for understanding when anthropomorphic system behaviors may or may not be desirable.

\noindent Finally, we believe that {\bf we also need to develop and use appropriate, precise terminology and language to describe anthropomorphic AI systems and their characteristics.} 
Discussions about anthropomorphic AI systems have regularly been plagued by claims of these systems attaining sentience and other human characteristics~\cite[e.g.,][]{chatbot-self-awareness,AI-self-awarness,AI-feelings,sentientGoogle}.
In line with existing concerns~\cite[e.g.,][]{cheng-etal-2024-anthroscore,dijkstra1985anthropomorphism,inie2024from,rehak2021language}, we believe that appropriately grounding and facilitating productive discussions about the characteristics or capabilities of anthropomorphic AI systems requires clear, precise terminology and language which does not carry over meanings from the human realm that are incompatible with AI systems. 
Such language can also help dispel speculative, scientifically unsupported portrayals of these systems, and support more factual descriptions of them.   

\bibliographystyle{plainnat}
\bibliography{references}
\end{document}